\documentclass[12pt]{iopart}
\usepackage{latexsym}
\def\R{{I\!\lhn\!R}}

\def\lhn{\;\!\!}
\def\Cl#1{{\cal C}\!\ell_{#1}}
\def\uls#1{{\scriptscriptstyle \underline{#1\!}\,}}
\def\ul#1{\underline#1}
\def\rCl#1{\mbox{\Large ${ }^{r}$}\!\Cl{#1}}
\def\rL{\mbox{\Large ${}^{r}$}\!\!\Lambda}
\def\sL{\mbox{\Large ${}^{s}$}\!\!\Lambda}
\def\rev{\widetilde}
\def\revg#1{(#1)\,\tilde{ }}
\def\conj#1{\overline{{#1}^{\!\raisebox{0.1cm}{\,}}} }
\def\conju#1{\,\,\overline{\!\! {#1}^{\!\raisebox{0.1cm}{\,}}\!}\, }
\def\main#1{{{#1}^*}}
\newcommand{\prj}[2]{{\langle #2 \rangle}_{#1}}
\newcommand{\prjs}[2]{{\langle \, #2  \, \rangle}_{#1}}

\def\f12{\frac{1}{2}}

\def\lit#1{\mbox{\raisebox{0.03cm}{${\scriptstyle #1}$}}}
\def\*{\! *\!}
\def\ot{\mbox{\raisebox{0.03cm}{${\scriptscriptstyle\bf\otimes}$}}}
\def\no{\noindent}

\def\w{\wedge}
\def\l{\lambda}

\def\beq{\begin{equation}}
\def\eeq{\end{equation}}
\def\beqna{\begin{eqnarray}}
\def\eeqna{\end{eqnarray}}
\def\beqa*{\begin{eqnarray*}}
\def\eeqa*{\end{eqnarray*}}
\def\inn{\!\cdot\!}
\def\L{\Lambda}
\def\o+{\oplus}

\def\tst{\textstyle}

\def\Spin{{\rm Spin}}
\def\Pin{{\rm Pin}}

\newtheorem{theorem}{Theorem}
\def\theore#1{\begin{theorem} #1$\Box$ \end{theorem}}
\def\prove#1{\noindent {\bf Proof.}  #1 \hfill \rule{2.5mm}{2.5mm}\vspace{3pt}}
\newtheorem{corollary}{Corollary}
\def\corollar#1{\begin{corollary} #1$\Box$ \end{corollary}}
\newtheorem{definition}{Definition}
\def\define#1{\begin{definition} #1$\Box$ \end{definition}}

\def\NCA{{\em Nuovo Cimento}}

\def\CQG{{\em Class.\ Quantum Grav.}}

\def\CMP{{\it Commun.\ Math.\ Phys. }}

\begin{document}

\title{Positivity and conservation of superenergy tensors}
\author{Jose M Pozo and Josep M Parra}
\address{Departament de F\'{\i}sica Fonamental,
Universitat de Barcelona\\
Diagonal, 647, E-08028 Barcelona, Spain}
\eads{\mailto{jpozo@ffn.ub.es}, \mailto{jmparra@ffn.ub.es}}

\begin{abstract}
Two essential properties of energy--momentum tensors $T_{\mu\nu}$ are their 
positivity and conservation. This is mathematically formalized by, 
respectively, an energy condition, as the dominant energy condition, and the 
vanishing of their divergence $\nabla^\mu T_{\mu\nu}=0$. The classical 
Bel and Bel-Robinson superenergy tensors, generated from the Riemann and 
Weyl tensors, respectively, are rank-4 tensors. But they share
these two properties with energy--momentum tensors: the {\it dominant property}
(DP) and the divergence-free property in the absence of sources (vacuum). 
Senovilla \cite{Seno1999a,Seno1999b} 
defined a universal algebraic construction which 
generates a {\it basic superenergy tensor} $T\{A\}$ from any arbitrary tensor 
$A$. In this construction, the seed tensor $A$ is structured as an $r$-fold 
multivector, which can always be done. The most important feature of the basic 
superenergy tensors is that they satisfy automatically the DP,
independently of the generating tensor $A$. In \cite{Pozo2000}, we presented a 
more compact definition of $T\{A\}$ using the $r$-fold Clifford 
algebra $\bigotimes^r\Cl{p,q}$. This form for the superenergy 
tensors allowed us to obtain an easy proof of the DP valid for any dimension.
In this paper we include this proof. We explain which new elements appear
when we consider the tensor $T\{A\}$ generated by a non-degree-defined 
$r$-fold multivector $A$ and how orthogonal Lorentz transformations and 
bilinear observables of spinor fields are included as particular cases of 
superenergy tensors. We find some sufficient conditions for the seed tensor 
$A$, which guarantee that the generated tensor $T\{A\}$ is divergence-free. 
These sufficient conditions are satisfied by some physical fields, which are 
presented as examples.
\end{abstract}

\pacno{0460}

\section{Introduction}

The source of gravitation in general relativity is the energy--momentum tensor
$T_{\mu\nu}$ of the matter content of the spacetime. The energy--momentum 
tensor of most physical fields satisfies (and is expected to satisfy) three 
well-known properties, which are a consequence of its definition from the 
fields involved and of the differential equations governing these fields. It 
is generally {\it symmetric}, $T_{\mu\nu}=T_{(\mu\nu)}$, and is 
{\it locally conserved}, i.e. divergence-free $\nabla^\mu T_{\mu\nu}=0$, in
the absence of sources or when we consider the total 
energy--momentum. These two properties are automatically satisfied by the 
Einstein tensor $S_{\mu\nu}$ of any pseudo-Riemannian manifold. Thus, they 
can also be seen as a consequence or requirement of Einstein equations 
$S_{\mu\nu}=8\pi GT_{\mu\nu}$. The third property of $T_{\mu\nu}$ is 
its `positivity'. For most fields, $T_{\mu\nu}$ satisfies the {\it dominant 
energy condition}, that is, the {\it energy flux} or {\it momentum vector}
\footnote{
	The sign is signature dependent. It is a minus sign in signature 
	($-$,+,+,$\cdots$) and a plus sign in the opposite.}
$j_\mu=-T_{\mu\nu}u^{\nu}$, 
measured by any future-pointing causal observer $u$, is also a 
future-pointing causal vector. When this is not satisfied, there arise 
problems of interpretation or rules of selection, as for the Tetrode tensor 
of a Dirac field. Indeed, this positivity condition (or another as
the {\it weak} or the {\it strong energy condition}), is usually required for 
the Einstein tensor of any physically acceptable spacetime.

The name superenergy was first applied to the Bel--Robinson
(BR) and Bel tensors \cite{Bel1962}, which are defined from the 
conformal Weyl tensor and
the Riemann tensor, respectively. The motivation for this name is
that they share some properties with energy--momentum tensors. 
The prefix super appears because they are rank-4 tensors instead 
of rank-2. 
In \cite{Seno1999a}, Senovilla defined an algebraic 
construction which generates a {\it basic superenergy tensor} $T\{A\}$, 
from an arbitrary {\it seed tensor} $A$. A much more extensive treatment 
discussing properties and applications is found in \cite{Seno1999b}. 
This construction unifies, in a single procedure, the BR and Bel 
tensors and many energy--momentum tensors from different physical 
fields. The most important feature of the basic superenergy tensors $T\{A\}$ 
is that, independently of the seed tensor $A$, they automatically satisfy 
the {\it dominant property} (DP), which is a generalization of the dominant 
energy condition for arbitrary rank tensors. This property holds in any 
dimension, provided that the space has Lorentzian signature, when the DP 
in fact can be defined.

An important application of these superenergy tensors has been the study of the 
interchange of {\it superenergy} between different fields, including matter 
content and gravitation \cite{Seno2000,Seno1999b}. 
The concept in mind is that the {\it total} 
superenergy may be conserved. This possibility would provide 
a physical relevance to the superenergy tensor of a field, in the same sense 
that the conservation of energy justifies even its physical `existence'. 
Success has been attained in some cases, in which a Killing vector exists. 
Independent of its physical significance, there are important mathematical 
applications of the basic superenergy tensor $T\{A\}$. The DP of the BR tensor 
is essential for the proof of the stability of Minkowski spacetime 
\cite{Christodoulou1993} 
and the fact that $T\{A\}$ can always be constructed has been used to provide a 
simple geometric criterion for the causal propagation of any physical field 
$A$. \cite{SenoBergq1999}

The original definition of $T\{A\}$ is formulated \cite{Seno1999b} in a 
standard index tensorial 
notation, but organizing tensors as $r$-fold exterior forms. This definition 
involves $2^r$ terms, corresponding to each combination of Hodge duals which 
can be obtained from an $r$-fold form. Here, we will use the name {\it $r$-fold 
multivector} instead of {\it $r$-fold form}. Since vectors and forms are 
considered identified by the metric, the difference in the name is simply a 
question of tradition.  
Bergqvist \cite{Bergqvist1999} reformulated $T\{A\}$ using 2-spinors.
The spinorial expression enabled an elegant proof of the DP.
However, this proof is only valid for four dimensions. A general proof for 
arbitrary dimension in tensorial notation was presented in 
\cite{Seno1999b}. 
In \cite{Pozo2000}, we presented an alternative formulation of $T\{A\}$ using
the Clifford algebra. Our expression is much more compact than both
standard-tensorial and spinorial expressions since it involves a unique term.
This was used to give a simple general proof of the DP for any dimension. 

In this paper we include the Clifford algebra formulation of superenergy 
tensors and the proof of the DP, presented in \cite{Pozo2000}. We especially 
emphasize that this expression is, in fact, a generalization of the $T\{A\}$ 
defined in \cite{Seno1999a,Seno1999b}, because it can be applied to 
non-degree-defined $r$-fold multivectors. In section \ref{sec:Symme}, we 
obtain the non-trivial new elements and properties which appear from this 
generalization. In section \ref{sec:Lorentz&Spinors}, we explain how orthogonal
Lorentz transformations and bilinear observables of spinors are included as 
particular cases of $T\{A\}$. A complete study of null cone preserving maps
has been recently presented in \cite{SenoBergq2001} where it was shown that
involutory Lorentz transformations are particular cases of superenergy tensors
generated by simple $n$-vectors. Here, we show that general Lorentz
transformations are also particular cases of superenergy tensors, those 
generated by the elements of the group Pin, which are in general inhomogeneous 
multivectors, being homogeneous $n$-vectors only for the case of involutory 
transformations.  

In section \ref{sec:PrinNulDir}, we introduce a general definition of the 
equation for a principal null direction $\ell$ of an arbitrary $r$-fold 
multivector $A$, 
which is the natural extension of the well-known cases for bivectors 
(electromagnetic field) and double bivectors (Weyl tensor). We prove
the equivalence of this equation with the vanishing of the generated 
superenergy tensor $T\{A\}$ when contracted $2r$ times with the null direction
$\ell$.

Section \ref{sec:Conserved} is concerned with the
conservation of $T\{A\}$ where we introduce some sufficient conditions on 
the seed tensor $A$ which guarantee that the basic superenergy tensor $T\{A\}$ 
generated by it, will be divergence-free. These sufficient conditions are 
implied by the differential equations satisfied by some fields, which we give 
as examples. These results might be useful to find conserved superenergy 
tensors combining different fields, since the conditions are easier to apply 
over the seed field $A$ than over the generated tensor $T\{A\}$. We consider 
that our results increase the mathematical relevance of the superenergy 
tensors, because any field $A$ of $r$-fold multivectors satisfying the 
sufficient conditions, automatically generates a tensor $T\{A\}$ which is 
conserved and satisfies the DP.

\section{Multivectors}
		\label{sec:Multivec}

Let us consider the tangent vector space ${\cal T}({\cal M})$ of some real 
manifold ${\cal M}$ of dimension $d=p+q$ endowed with a metric $g$ of
signature $p-q$. Via the exterior product, it generates the {\it Grassmann} or 
{\it exterior space} $\L\equiv\L({\cal T}({\cal M}))$, which is a graded space 
$\L=\L_0\o+\L_1\o+\cdots\o+\L_d$, where $\L_0$ contains the scalars, $\L_1$ 
the vectors, $\L_2$ the bivectors and in general $\L_i$ contains the 
$i$-vectors. A general element of the Grassmann space $A\in\L$ is called a
{\it multivector}, and it is said to be {\it homogeneous} if it 
is an $i$-vector for some degree $i$, $A\in\L_i$; otherwise it is said to be
{\it inhomogeneous}. 
A basic operation on multivectors is the {\it degree projection}, which
projects into some subspace $\L_i$. This is denoted by angle brackets:
\[
\forall A\in\L \qquad
A=\sum_{i=0}^d \prj{i}{A} \qquad \mbox{where} \quad \prj{i}{A}\in\L_i .
\]
To write a multivector $A\in\L$ in a complete basis of 
the linear space $\L$, we will use a {\it multi-index}, denoted
by a Latin capital letter,
\[ 
	A=A^Ie_{I} .
\]

Besides the exterior product, which is metric-independent, the metric $g$ 
defines another associative \cite{Riesz1958} product on multivectors, called 
the {\it Clifford geometric product} \cite{Hestenes1985,Graf1978,Kahler1962}. 
We will denote the Clifford
product by simple juxtaposition. As a simple case, the product of a general
multivector $A\in\L$ with a vector $b\in\L_1$ can be expanded into inner and
exterior products:
\[
	Ab=A\inn b+A\w b \qquad {\rm and} \qquad bA=b\inn A+b\w A  .
\] 
The Grassmann space $\L$ endowed with this product constitutes the 
{\it Clifford geometric algebra} $\Cl{p,q}$. 

There are one natural linear involution and two linear anti-involutions, which 
are independent of the metric but are fixed by the graded structure of $\L$: 
The {\it main involution}, denoted by an asterisk $\main{A}$, is defined as
the involution, 
$\main{(AB)}=\main A\main B\quad \forall A,B\in\L$, 
which changes each vector into its opposite $\main{a}=-a\quad\forall a\in\L_1$.
The {\it reversion}, denoted by a tilde $\rev{A}$, is defined as the 
anti-involution, 
$\rev{\! AB}=\rev{B}\rev{A}\quad\forall A,B\in\L$, 
which keeps vectors unchanged $\rev{a}=a\quad\forall a\in\L_1$.
The {\it Clifford con\-ju\-ga\-tion}, denoted by 
an overline $\conju{A}$, is the composition of the reversion and the main
involution, $\conju{A}=\main{\rev A}$. Taking, for instance,
a factorizable trivector
\[ 		\fl
	\revg{a\w b\w c}=c\w b\w a=-a\w b\w c \qquad
	\conj{a\w b\w c}=(-c)\w(-b)\w(-a)=a\w b\w c  .
\]

\section{${\bf r}$-fold multivectors}
		\label{sec:rFoldMultivec}

An {\it $r$-fold multivector} A is a rank-$r$ tensor on the Grassmann space 
$\L$
\[ 
	A\in \rL \equiv \bigotimes^r \L = T^r_0(\L)  .
\]

\no Each Grassmann space $\L$ will be called a {\it
block}. Its expression in a basis, using multiindices, is
\beq \label{r-fold}
	A=A^{I_1 \, I_2 \ \cdots \ I_r} \,
	e_{I_1}\ot\, e_{I_2}\ot\,\cdots\,\ot\, e_{I_r}  .
\eeq
We observe that terms of the form $e_{I_1}\ot 1\ot e_{I_3}$ are possible and
are different from $e_{I_1}\ot e_{I_3}$.

A natural associative product defined between two $r$-fold 
multivectors is the {\it $r$-fold Clifford product}, which involves 
an independent Clifford product in each block: 
\beqa*
	A B&=&\big( A^{I_1 \,\cdots\, I_r} \,e_{I_1}\ot
	\cdots\ot\, e_{I_r}\big)\,\big(B^{J_1 \,\cdots\, J_r}
	\, e_{J_1}\ot\cdots\ot\, e_{J_r}\big) 
\\
	&=&A^{I_1 \,\cdots\, I_r}B^{J_1 \,\cdots\, J_r}\,(e_{I_1}e_{J_1})
	\,\ot\,\cdots\,\ot\, (e_{I_r}e_{J_r})  .
\eeqa*
The space $\rL$ endowed with this product constitutes the {\it $r$-fold
Clifford algebra}: 
$\rCl{p,q}\equiv \bigotimes^r \Cl{p,q}$.
We will also consider the product between an $r$-fold and an $s$-fold
multivector with $r\neq s$. This is defined by an independent Clifford
product in each block, starting from the left-hand side. If $A\in\rL$ and 
$B\in\sL$ where, say, $s<r$ then
\[
	AB=A^{I_1 \,\cdots\, I_r}B^{J_1 \,\cdots\, J_s}\,
	(e_{I_1}e_{J_1})\,\ot\,\cdots\,\ot\, (e_{I_s}e_{J_s})\,\ot 
	\,e_{I_{s+1}}\,\ot\,\cdots\,\ot\,e_{I_r}  .
\]
Note that this product is equivalent to the product with $r=s$, if 
we complete the right-hand side of the shortest factor with the 
necessary number of `ones'
\[
	\sL\hookrightarrow\rL\qquad 
	B^{J_1 \,\cdots\, J_s}\,e_{J_1}\ot\cdots\ot\, e_{J_s}
	\ \mapsto\ 
	B^{J_1 \,\cdots\, J_s}\,e_{J_1}\ot\cdots\ot\, e_{J_s}
	\,\ot\,1\,\ot\,\cdots\,\ot\,1 
\]
which defines the {\it canonical $s$$\hookrightarrow$$r$ left-immersion}.

In order to shorten expressions, we introduce {\it multi-fold multi-indices}.
They collect a list of multiindices and are denoted by an 
underline, $ \ul{I}\equiv\{I_1,I_2,\cdots ,I_s\} $.
Using them, expression (\ref{r-fold}) simplifies to
$ A=A^{\,\uls{I}} \, e_{\uls{I}} $.
They will also be used as a shorthand to make a
single block explicit:
\[
	A=A^{\uls{C}\, I_i\, \uls{D}} \, 
	e_{\uls{C}}\,\ot\, e_{I_i}\ot\, e_{\uls{D}} =
	e_{\uls{C}}\,\ot \;\! A^{\uls{C}\,\uls{D}}\,\ot\, e_{\uls{D}}
\]

\no where $\ul{C}=\{I_1,\cdots ,I_{i-1}\}$ and  
$\ul{D}=\{I_{i+1},\cdots ,I_r\}$. Note that 
$A^{\uls{C}\,\uls{D}}\equiv A^{\uls{C}\, I_i\, \uls{D}}\, e_{I_i}\in\L$
is not just a scalar component but a multivector.

The basic operations of degree projection, reversion and Clifford 
conjugation, acting on multivectors, can be extended to $r$-fold
multivectors. Thus, we define the {\it $r$-fold degree projection},
the {\it $r$-fold reversion} and the {\it $r$-fold Clifford 
conjugation} as the result of applying the corresponding
operation independently to every block, and will use the same
notation as with simple multivectors:
\[ 		\fl
	\prj{s_1,s_2,\cdots,s_r}{A}=A^{I_1 \, I_2 \, \cdots \, I_r} 
	\prj{s_1}{e_{I_1}}\ot \prj{s_2}{e_{I_2}}\ot\,\cdots\,\ot 
	\prj{s_r}{e_{I_r}} \quad\qquad
	\prj{s}{A}\equiv\prj{s,s,\cdots,s}{A}
\]
\[		\fl
	\rev{A}=A^{I_1 \, I_2 \,\cdots\, I_r}\,\rev{e_{I_1}}\ot
	\,\rev{e_{I_2}}\ot\,\cdots\,\ot\,\rev{e_{I_r}}
	\qquad \mbox{and} \qquad
	\conju{A}=A^{I_1 \, I_2 \,\cdots\, I_r}\,\conj{e_{I_1}}\ot
	\,\conj{e_{I_2}}\ot\,\cdots\,\ot\,\conj{e_{I_r}}  .
\]

As remarked by Senovilla \cite{Seno1999b}, any tensor 
$\hat{A}=\hat{A}^{\mu_1 \,\cdots\, \mu_s}\,e_{\mu_1} 
\ot\,\cdots\,\ot\, e_{\mu_s}$ can be considered as an $r$-fold multivector,
by reordering and grouping antisymmetric indices in separated blocks.
Thus, the {\it reordered tensor} $A$ will be an
{\it $r$-fold ($n_1,n_2,\ldots,n_r$)-vector},
\[
	A\,\in\,\L_{n_1}\ot\L_{n_2}\ot\cdots\ot\L_{n_r}
	\, \subset \,  
	\rL
\]
where $n_1+n_2+\cdots+n_r=s$. An $r$-fold multivector $A$ which is homogeneous 
at each block is said to be {\it degree-defined}. 

\section{Superenergy tensors}
		\label{sec:SEtensors}

The procedure for the construction of the {\it basic superenergy tensor} 
\cite{Seno1999b} $T\{A\}$ starts with the arrangement of the seed tensor $A$ 
as an $r$-fold multivector, in a way that has been commented above. 
Strictly speaking, we will consider that $A$ is a seed $r$-fold multivector.
The procedure was originally defined for the signature $p-1$ ($d=p+1$), while 
its expression in 2-spinors language, obtained by Bergqvist 
\cite{Bergqvist1999}, is naturally written for signature $1-p$. 
Here, we will introduce the Clifford algebra formalism expression for the 
signature $p-1$, that is, for the algebra $\Cl{p,1}$, but 
the corresponding expression for the opposite signature is also given.

Senovilla's definition of the basic superenergy tensor has the following form 
\cite{Seno1999b}:
\beq 
		\label{Tens_def}
	T\{A\}=\f12 \sum_{\cal P} A_{\cal P}\times A_{\cal P} 
\eeq

\no where $A_{\cal P}$ denotes the $r$-fold multivector $A$
transformed by a combination of duals: that is, ${\cal P}$ codifies the action
of taking the Hodge dual on some blocks and keeping the rest of blocks
unchanged. Thus, the summation runs through the $2^r$ possible 
combinations. The cross product $A\times A$ used in (\ref{Tens_def}) is defined
as the contraction in every block of all the indices except one in each 
factor. Therefore, $T\{A\}$ has $r$ pairs of indices:
\beq		\fl
		\label{cross}
	(\!A\!\times\! A)_{\mu_1 \nu_1 \ \cdots \ \mu_r \nu_r}\!=\! 
	\left(\,\prod_{i=1}^r
	\frac{1}{(n_i-1)!}\right)
	A_{\mu_1 \,\l_{12}\cdots \l_{1n_1}\ \cdots \  
	\mu_r \,\l_{r2}\cdots \l_{rn_r}}
	\ A_{\nu_1}{}^{\l_{12}\cdots \l_{1n_1}}
	{}^{\ \cdots}_{\ \cdots \ \nu_r}{}^{\l_{r2}\cdots \l_{rn_r}} .
\eeq

The basic superenergy tensor $T\{A\}$ is a generalization of some well-known 
superenergy and energy--momentum tensors. A number of examples are given in 
\cite{Seno1999b}. Among them are the Weyl conformal tensor $C$ which generates 
the Bel--Robinson superenergy tensor $T\{C\}$, the Riemann tensor $R$ which
generates the Bel tensor $T\{R\}$, the gradient of a Klein--Gordon massless 
scalar field $\nabla\phi$ which generates its standard energy--momentum tensor 
$T\{\nabla\phi\}$ and, the Faraday bivector $F$ which generates the standard 
electromagnetic energy--momentum tensor $T\{F\}$. We refer the reader to 
section \ref{sec:Conserved} and to the bibliography for more details.

The use of the $r$-fold Clifford algebra $\rCl{p,1}$ enables us to introduce 
an alternative and much more compact definition of the basic superenergy 
tensor. Our expression is inspired by the Clifford geometric algebra 
formulation of the standard electromagnetic energy--momentum tensor 
\cite{Hestenes1966} which defines this tensor as a vector endomorphism:
\[
	T(u)=-\f12 F\,u\,\conju{F} \in\L_1 \qquad \forall u\in\L_1 .
\]
Equivalently, one can write it as applied to a pair of vectors, or 
obtain its components in a basis by simply applying the tensor to the elements 
of this basis (first found in this form in \cite{Riesz1946}):
\[
	T(v,u)=-\f12 \prj0{v Fu\conju{F}} \qquad{\rm and}\qquad
	T_{\mu\nu}=-\f12\prj0{e_\mu Fe_\nu \conju{F}} .
\]
This kind of sandwich-like formula is really a standard procedure in Clifford
algebra formulations which, as we will see below, also appears in the 
implementation of isometries and for obtaining the bilinear observables of a 
spinor field.

Generalizing this expression for $r$-fold multivectors $A$, we define the 
associate superenergy tensor as an endomorphism on the direct 
product of $r$ copies of the vector space, 
\[
	T\{A\}:{\tst \bigotimes^r\L_1 \rightarrow \bigotimes^r \L_1}\qquad
	u\mapsto T\{A\}(u):=(-1)^r\f12\prj1{A\, u\,\conju{A}\,} .
\]
Thus, applied to $r$ vectors,
\beq
	T\{A\}(u_1\ot u_2\ot\cdots\ot u_r) \ = \ 
	(-1)^r\f12\prj1{A \, (u_1\ot\cdots\ot u_r) \,\conju{A}\,}
		\label{Clif_def0}  .
\eeq 
We can also write it as applied to $r$ pairs of vectors
\beq		\fl
	T\{A\}(v_1\ot\cdots\ot v_r,\,u_1\ot\cdots\ot u_r) \ = \ (-1)^r \f12
	\prj0{(v_1\ot\cdots\ot v_r)\,A \,(u_1\ot\cdots\ot u_r) \,\conju{A}\,}
		\label{Clif_def}
\eeq 
or obtain its components in a basis $\{e_\mu\}$
\beq
		\label{Components}
	{T\{A\}}_{\mu_1\,\nu_1 \ \cdots \ \mu_r\,\nu_r}=(-1)^r\f12
	\prj0{(e_{\mu_1}\ot\cdots\ot\, e_{\mu_r})\,A \ 
	(e_{\nu_1}\ot\cdots\ot\, e_{\nu_r})\,\conju{A}\, }  .
\eeq

The expression of the superenergy tensor for the signature $1-p$ is
slightly more simple: $T\{A\}(u)=\f12\prj1{Au\rev{A}}$. Indeed, the reason for
the sign $(-1)^r$ is that, for the signature $p-1$, a time-like vector must be 
{\it past-pointing} to give a positive quantity when contracted with a 
{\it future-pointing} vector.

In the remaining part of the section, we will show that both expressions, 
Senovilla's standard-tensorial definition (\ref{Tens_def}) and the 
Clifford algebra formulation (\ref{Clif_def}), correspond to the same object. 
The method for proving this identity will be expansion of (\ref{Clif_def}) in 
order to obtain (\ref{Tens_def}). We must remark here that Senovilla's 
definition applies to degree-defined $r$-fold multivectors. In contrast, 
expression (\ref{Clif_def}) can be applied to general $r$-fold multivectors. We
will comment on this generalization later. But evidently, for this proof, we 
must consider only degree-defined $r$-fold multivectors.

Let us concentrate on a single arbitrary block. 
Using multi-fold multi-indices, the components (\ref{Components})
can be written as 
\[ 
	T\{A\}_{\mu_1 \nu_1 \cdots \mu_r \nu_r}= 
\]
\[ 		\fl
	(-1)^r\f12
	\prj0{e_{\mu_1,\cdots,\mu_{i-1}}e_{\uls{C}}
	e_{\nu_1,\cdots,\nu_{i-1}}\conj{e_{\uls{E}}}}
	\,\prj0{e_{\mu_i}\, A^{\uls{C}\uls{D}}\ 
	e_{\nu_i}\, \conju{A^{\uls{E}\uls{F}}}}
	\,\prj0{e_{\mu_{i+1},\cdots,\mu_r}e_{\uls{D}}
	e_{\nu_{i+1},\cdots,\nu_r}\conj{e_{\uls{F}}}} .
\]

To re-express the result in this $i$th block, we take into account
two essential facts. The first
is that the Clifford product of any multivector with a vector can be 
split into inner and exterior products:
\[
e_{\mu_i}\, A^{\uls{C} \uls{D}}=
e_{\mu_i}\cdot A^{\uls{C} \uls{D}}+
e_{\mu_i}\w A^{\uls{C} \uls{D}} .
\]
With this expansion the block
$\prj0{(e_{\mu_i} A^{\uls{C}\uls{D}})\ 
(e_{\nu_i} \conju{A^{\uls{E}\uls{F}}})}$
should split into four terms. But since $A$ is degree-defined, the cross terms 
cannot contract completely and vanish. Thus, it splits into two terms,
\beq		\fl
		\label{crossTerms}
	\prj0{e_{\mu_i}\,A^{\uls{C} \uls{D}}\
	e_{\nu_i}\,\conju{A^{\uls{E} \uls{F}}}}=
	\prj0{ ( e_{\mu_i}\cdot A^{\uls{C} \uls{D}})\,
	( e_{\nu_i}\cdot \conju{A^{\uls{E} \uls{F}}}) }
	+ \prj0{ ( e_{\mu_i}\w A^{\uls{C} \uls{D}})\,
	( e_{\nu_i}\w \conju{A^{\uls{E} \uls{F}}}) }  .
\eeq

The second fact is that an exterior product can be written, with
the help of the Hodge duality, as an inner 
product. Applying it, the second term has the same structure as
the first term, but whereas in the first 
we have the original $n_i$-vector, $A^{\uls{C} \uls{D}}$, in the second
we have the dual $(d\!-\!n_i)$-vector, $*A^{\uls{C} \uls{D}}$,
\[		\fl
	\prj0{e_{\mu_i}\,A^{\uls{C} \uls{D}}\
	e_{\nu_i}\,\conju{A^{\uls{E} \uls{F}}}}=
	\prjs0{ ( e_{\mu_i}\cdot A^{\uls{C} \uls{D}})\,
	( e_{\nu_i}\cdot \conju{A^{\uls{E} \uls{F}}})  }+
	\prjs0{  \big( e_{\mu_i}\cdot [*A^{\uls{C} \uls{D}}]\big)\,
	\big( e_{\nu_i}\cdot \conju{[*A^{\uls{E} \uls{F}}]\!}\big)  } .
\]

Repeating this expansion for every block we obtain $2^r$
terms, corresponding to all possible combinations that take 
the dual in some blocks and leave the rest unchanged,
\beq		\fl
		\label{Summation}
	T\{A\}_{\mu_1 \nu_1 \cdots \mu_r \nu_r}=
	(-1)^r\f12\sum_{\cal P}
	\prjs0{ ( \lit(e_{\mu_1}\ot\cdots\ot\,e_{\mu_r}\lit) \cdot A_{\cal P} ) 
	\ ( \lit(e_{\nu_1}\ot\cdots\ot\, e_{\nu_r}\lit) \cdot 
	\conju{A_{\cal P}} ) }
\eeq

\no where the dot denotes the inner product in every block, and we have
used ${\cal P}$ again to indicate each combination of Hodge duals.
Finally, comparing this last expression with (\ref{Tens_def}), we only have
to check that the terms of this summation (\ref{Summation}) coincide with 
Senovilla's cross product (\ref{cross}),
\[		\fl
	(-1)^r\prj0{(e_{\mu_1,\cdots,\mu_r}\cdot A)\,
	(e_{\nu_1,\cdots,\nu_r}\cdot\conju{A})}\, = \,
	\prj0{(e_{\mu_1,\cdots,\mu_r}\cdot A)\,
	\rev{(e_{\nu_1,\cdots,\nu_r}\cdot A)\raisebox{4mm}{ }}\!\!}
	\, = \, (A\times A\,)_{\mu_1\nu_1\cdots\mu_r\nu_r} .
\]
This can be seen by realizing that the dot products in (\ref{Summation})
fix one index in each factor for each block. The scalar
projection of the product selects the terms that correspond to
the contraction of the rest of the indices. It is easy to 
check that the signs coincide. Then, the proof is complete.

\section{Symmetries}
		\label{sec:Symme}

The basic superenergy tensor defined in \cite{Seno1999b} is symmetric, by
construction, in each pair of indices:
\[
	T\{A\}_{\mu_1\nu_1\mu_2\nu_2\cdots\mu_r\nu_r}=
	T\{A\}_{(\mu_1\nu_1)(\mu_2\nu_2)\cdots(\mu_r\nu_r)} .
\]
However, this property only holds for a degree-defined $r$-fold 
multivector $A$. When $A$ is non-degree-defined, formula 
(\ref{crossTerms}) is no longer valid because there can be contributions 
from terms mixing different degrees. Let us
see which non-trivial elements appear when we consider general $r$-fold
multivectors.

First, consider a simple multivector as a seed, split it into its different 
degrees $A=\sum_{k=0}^d A_{[k]}$, where $A_{[k]}\in\L_k$. Then, the 
contribution of each degree to the tensor T\{A\} is given by
\beqa*		\fl
	- 2T\{A\}(u,v) &=&
	\sum_{k=0}^{d}\prj0{\prj{k}{uA}\prj{k}{v\conju{A}}}
\\
	&=&\sum_{k}\prjs0{(u\inn A_{[k+1]}+u\w A_{[k-1]})\,
	(v\inn \conju{A_{[k+1]}\!}\, + v\w \conju{A_{[k-1]}\!}\,)}
\\
	&=&\sum_k\ \ \prj0{(u\inn A_{[k+1]})(v\inn \conju{A_{[k+1]}\!}\,)}+
	\prj0{(u\w A_{[k-1]})(v\w \conju{A_{[k-1]}\!}\,)}
\\
	&& + \prj0{(u\inn A_{[k+1]})(v\w \conju{A_{[k-1]}\!}\,)}+
	\prj0{(u\w A_{[k-1]})(v\inn \conju{A_{[k+1]}\!}\,)}  .
\eeqa*
The first pair of terms was already present in formula (\ref{crossTerms}), 
while the last pair collects the non-vanishing terms mixing different degrees. 
Observe that these extra terms only contain factors differing in two degrees.
This means, for instance, that a bivector combines only with scalars and 
4-vectors. Besides, the contribution of these terms is always antisymmetric in
the pair of indices. Then, the total superenergy tensor $T\{A\}$ can be written
as the sum of the symmetric superenergy tensors generated by each degree and 
the antisymmetric mixing terms:
\[
	T\{A\}(u,v)=\sum_k T\{A_{[k]}\}(u,v) + 
	\sum_k\prj0{\prj2{A_{[k+2]}\rev{A_{[k]}}}(v\w u)}
\]
or written with tensorial indices
\[
	T\{A\}_{\mu\nu}=\sum_k T\{A_{[k]}\}_{\mu\nu} + 
	\sum_k \frac{1}{k!}(A_{[k+2]})_{\mu\nu\rho_1\cdots\rho_k}
		(A_{[k]})^{\rho_1\cdots\rho_k}  .
\]
For the more general case of $r$-fold multivectors, this splitting will occur
in each block. Thus, it can appear as any combination of terms symmetric in 
some blocks and antisymmetric in the rest, so that no symmetry is satisfied in 
general. 

Another possible symmetry is the one involving the interchange of different 
pairs of indices. It is easy to see that, if $A\in\rL$ is symmetric or 
antisymmetric in any two blocks, then, $T\{A\}$ will be symmetric in the
interchange of the respective pair of indices:
\[		\fl
	A^{I_1\cdots I_i\cdots I_j\cdots I_r}=
	\pm A^{I_1\cdots I_j\cdots I_i\cdots I_r}\quad\Rightarrow\quad
	T\{A\}_{\mu_1\nu_1\cdots\mu_i\nu_i\cdots\mu_j\nu_j\cdots\mu_r\nu_r}=
	T\{A\}_{\mu_1\nu_1\cdots\mu_j\nu_j\cdots\mu_i\nu_i\cdots\mu_r\nu_r}  .
\]
This property is independent of $A$ being degree-defined or not. 

\section{Lorentz transformations and bilinear observables of spinors}
		\label{sec:Lorentz&Spinors}

A general Lorentz transformation of a vector $u\in\L_1$ is performed by a 
{\it versor} $R\in\Pin_{p,1}\subset\Cl{p,1}$, by means of the formula 
\cite{Crumeyrolle1990},
\beq
		\label{Lorentz}
	{\cal R}(u)=R u \main{R}^{-1} .
\eeq
The group $\Pin_{p,1}$ is the double covering group of O$(p,1)$, and is 
defined by
\[
	\Pin_{p,1}\equiv \{\ R\in\Cl{p,1} \ |\  R\rev{R}=\pm1
	\quad{\rm and}\quad 
	R u \main{R}^{-1}\in\L_1\ \ \forall u\in\L_1\ \} .
\]
By means of the Cartan--Dieudonn\'e theorem it is shown \cite{Crumeyrolle1990} 
that $\Pin_{p,1}$ can be equivalently defined by
\beq		\fl
		\label{PinFactors}
	\Pin_{p,1}\equiv \{\ R=v_1 v_2\cdots v_n\in\Cl{p,1} \ |\ 
	v_i\in\L_1,\ \ {v_i}^2=\pm1\quad{\rm and}\quad
	n=0,1,\cdots,2d \ \} .
\eeq 

The sign $R\rev{R}=\pm1$ splits the group into two unconnected parts, 
$\Pin^\pm_{p,1}$, which implement, respectively, {\it orthochronous} and 
{\it time-reversing} Lorentz transformations. Taking into account this 
sign to compute the inverse in expression (\ref{Lorentz}), we obtain
\[
	{\cal R}(u)=\pm R u \conju{R}
	\qquad {\rm when} \quad R\in\Pin^\pm_{p,1} .
\]
This can be seen as a particular case of the basic superenergy tensor
$T\{R\}$ when $R\in\Pin_{p,1}$. 
The opposite of a time-reversing transformation, $-{\cal R}(u)$, 
is an orthochronous transformation. This implies that 
$Ru\conju R=\mp{\cal R}(u)$, $R\in\Pin_{p,1}^\pm$ is always orthochronous. 
This proves the following:

\theore{	\label{LorentzMap}
A mapping $L:\L_1\rightarrow\L_1$ is an orthochronous Lorentz 
transformation iff there exists some $R\in\Pin_{p,1}$ for which
$L(u)=-2T\{R\}(u)=Ru\conju{R}$. }

This theorem together with the expression (\ref{PinFactors}) for the Pin
group can be considered as a generalization of corollary 4.2 in 
\cite{SenoBergq2001}. This corollary states that a superenergy tensor 
generated by a non-null simple $n$-vector $A=v_1\w v_2\w \cdots\w v_n$ 
defines a mapping
\footnote{
	The minus sign is due to the signature $p\!-\!1$.
	This is not present in \cite{SenoBergq2001}, which uses
	signature $1\!-\!p$. }
$-T\{A\}(u)$ proportional to an involutory orthochronous Lorentz 
transformation.

Observe that, in general, the versor $R\in\Pin_{p,1}$ in theorem 
\ref{LorentzMap} is not a homogeneous multivector.
This is why general orthochronous Lorentz transformations are not
included as superenergy tensors in \cite{SenoBergq2001}. 
$R\in\Pin_{p,1}$ is homogeneous only when it can be decomposed into 
orthogonal vectors:
\[
	R=v_1 v_2\cdots v_n=v_1\w v_2\w \cdots\w v_n .
\]
In this case, $R$ is a simple $n$-vector and implements an involutory 
transformation, which is implied by the property $R^2=\pm1$.

The special case of orthochronous {\it proper} Lorentz transformation is 
performed by a {\it rotor} 
$R\in\Spin_{p,1}^+\equiv\Pin^+_{p,1}\cap \Cl{p,1}^+$. The 
group $\Spin_{p,1}^+$ is the connected double covering group
of SO$^+(p,1)$, and $\Cl{p,1}^+$ is the even subalgebra of $\Cl{p,1}$.

Let us now consider an $r$-fold 1-vector $\ul{u}\in\bigotimes^r\L_1$. 
Then, we can simultaneously implement an independent orthochronous Lorentz 
transformation at each block, by means of an {\it $r$-fold versor}
\beq		\fl
		\label{r-foldRotations}
	\ul{R}=R_1\ot R_2\ot\cdots\ot R_r \qquad R_i\in\Pin_{p,q}^+
	\qquad {\cal R}(\ul{u})=\ul{R}\,\ul{u}\,\conju{\ul{R}}=
	(-1)^r \,2T\{\ul{R}\}(\ul{u}) .
\eeq

We can extend the group of Lorentz transformations to include dilations. 
The simplest way of implementing these transformations is to extend the 
$\Pin_{p,1}^+$ group by adding a scalar factor:
\[
	\Gamma_{p,1}^+\equiv
	\{\, \l R\ |\ \l\in\R^+,\  R\in\Pin_{p,1}^+\, \}
	\qquad u\mapsto (\l R)u\conju{(\l R)}=\l^2 {\cal R}(u) .
\]
$\Gamma_{p,1}^+$ is the {\it positive} subgroup of the 
Clifford--Lipschitz group \cite{Crumeyrolle1990}
\beqa*
	\Gamma_{p,1}&\equiv& \{\ R\in\Cl{p,1} \ |\  
	R u \main{R}^{-1}\in\L_1\ \ \forall u\in\L_1\ \}  \\
	&=&\{\ R\in\Cl{p,1} \ |\  Ru\conju{R}\in\L_1\ \ \forall u\in\L_1
	\quad{\rm and}\quad
	R\rev{R}\in\R^*\ \} .
\eeqa*
The extension for the transformation of $r$-fold 1-vectors is completely
analogous to (\ref{r-foldRotations}).

The singular limiting set
\[
	\Gamma_{p,1}^0\equiv\{\ R\in\Cl{p,1} \ |\  
	Ru\conju{R}\in\L_1\ \ \forall u\in\L_1 
	\quad{\rm and}\quad R\rev{R}=0\ \}
\]
is a semigroup which contains non-invertible even multivectors.
Each element $R\in\Gamma_{p,1}^0$ maps any vector $u$ into a unique
null vector $\ell=Ru\conju{R}$, up to a scalar factor. Thus, it defines
a null direction.

Pauli, Dirac, Weyl and Majorana spinor fields can be treated as {\it operator
spinors} \cite{Hestenes1973,Lounesto1996}. 
The concept of operator spinor is based on 
identifying spinors as elements of the even subalgebra $\Cl{p,q}^+$. 
The name is motivated because a spinor is then interpreted as a generalization
of a rotor, which links the observer tetrad field $\{e_\mu\}$ to 
the bilinear observables associated with the spinor. 
Typically, for a Dirac spinor $\Psi\in\Cl{3,1}^+$, the current and the spin
vectors, and the magnetization bivector are given by
\[
	j=\Psi e_0\conj{\Psi}\quad,\qquad s=\Psi e_3 \conj{\Psi}
	\qquad{\rm and}\qquad M=\Psi e_{12}\conj{\Psi} .
\] 
For Weyl and Majorana spinors, these expressions are also valid. However, 
these spinors are not general even multivectors but elements of two specific 
subsets of $\Gamma_{p,1}^0\cap\Cl{p,1}^+$. Thus, their current $j$ 
is light-like.

Operator spinors are represented by multivectors. But they are not
{\it tensors} because they relate the observables, which are objective 
quantities (tensors), to the observer. They can be 
considered `pseudo'-tensors in the same sense as affine connections. Thus, a 
spinor is inseparably related to the observer tetrad. For this reason, its 
associated superenergy tensor $T\{\Psi\}_{\mu\nu}$, is only a tensor for
the first index, since the second is related to the observer: 
$j_\mu=-2T\{\Psi\}_{\mu\nu}(e_0)^\nu\equiv -2T\{\Psi\}_{\mu(0)}$.
Thus, while a tensorial multivector A generates a rank-2 tensor $T\{A\}$, a
spinor $\Psi$ generates a rank-1 tensor $T\{\Psi\}(e_0)$. Compare this 
treatment with the 2-spinor formulation in \cite{SenoBergq1999}.

\section{Dominant property (DP)}
		\label{sec:DP}

In this section, we present a simple proof of the DP for the superenergy tensor 
$T\{A\}$, using its expression in the $r$-fold Clifford algebra $\rCl{1,p}$. 
We must underline that this proof is valid for a general $r$-fold multivector 
$A$. Even if the seed tensor $A$ is degree-defined, the essential point of 
the present proof is the definition of a second $r$-fold multivector $A'$, 
which in general is no longer degree-defined. We see, then, that this 
generalization is fundamental for the proof.

\define{A superenergy tensor satisfies the 
DP if for all collection $\{u_i,v_i\}$ of causal 
and future-pointing (f-p) vectors 
\[
	T\{A\}(u_1\ot\cdots\ot u_r,\,v_1\ot\cdots\ot v_r)=
	\f12\prjs0{(u_1\ot\cdots\ot u_r) \,A \, 
	(v_1\ot\cdots\ot v_r) \,\conju{A} }\geq 0 .
\]
}

\theore{Let $A\in\rL$ be any $r$-fold multivector, then $T\{A\}$ satisfies the
DP. }

\prove{{Let us recall, first, that a time-like f-p vector $u$ can always 
be expressed as the result of applying a local orthochronous Lorentz 
transformation and a dilation to a chosen unitary time-like f-p vector $e_0$.
As seen in the previous section, this transformation is performed by means of 
a multivector:
\[ 
	u=R_u e_0 \conju{R_u} \qquad 
	R_u\in\Gamma_{p,1}^+\subset \Cl{p,1} .
\]
The same expression applies for a null vector $u$ with 
$R_u\in\Gamma_{p,1}^0$. For the tensor product of $r$ f-p vectors
$\ul{u}\equiv u_1\ot u_2 \ot\cdots\ot u_r\in\bigotimes^r\L_1$, we can construct
\[		\fl
	R_{\ul{u}}\equiv R_{u_1}\ot R_{u_2}\ot\cdots\ot R_{u_r}  
	\quad{\rm and}\quad
	\ul{e_0}\equiv\underbrace{e_0\ot\cdots\ot e_0}_r
	\quad\mbox{so that}\quad 
	\ul{u}=R_{\ul{u}} \ \ul{e_0} \,\conju{R_{\ul{u}}} .
\]
\indent Our proof of the DP proceeds in two steps. First, using the operators 
$R_{\ul{u}}\in\rCl{p,1}$, we express
the result of applying $T\{A\}$ to any set of $2r$ f-p vectors 
$\ul{u},\ul{v}$, as the {\small $\{0,\ldots,0\}$} component of another 
superenergy tensor $T\{A'\}$:
\beqna		\fl
			\label{Aprime}
	(-1)^r\,2 T\{A\}(\ul{u},\ul{v})&=&
	\prjs0{\ul{u}\, A \, \ul{v} \,\conju{A}}
	=
	\prjs0{\big(R_{\ul{u}} \ \ul{e_0} \,\conju{R_{\ul{u}}}\big) \, A
	\,\big(R_{\ul{v}} \ \ul{e_0} \,\conju{R_{\ul{v}}}\big)\, \conju{A}\, }\\
			\nonumber
	&=&
	\prjs0{\ul{e_0}\ \big( \conju{R_{\ul{u}}} A R_{\ul{v}}\big) \ 
	\ul{e_0}\ \conju{ \:\!\big( \conju{R_{\ul{u}}} A R_{\ul{v}} \big) \!} }
	=\ 
	(-1)^r\,2 T\{A'\}(\ul{e}_0,\ul{e}_0)
\eeqna
where $A'\equiv\conju{R_{\ul{u}}} A R_{\ul{v}} \in\rCl{p,1}$ 
is also an $r$-fold multivector. \\
\indent The second step proves that, for all $A'\in\rCl{p,1}$, this component 
is non-negative:
\[
	T\{A'\}(\ul{e}_0,\ul{e}_0)=
	(-1)^r \f12 \prjs0{\ul{e_0}\ A'\ \ul{e_0}\,\conju{A'}}
	\geq 0 \qquad \forall A'\in\rCl{p,1} .
\]
To see this, we split $A'$ into parts orthogonal 
and parallel to the direction $e_0$. This splitting corresponds to the 
isomorphism of linear spaces, though not as algebras,
\[
	\rCl{p,1}\simeq\rCl{p,0}\otimes\rCl{0,1}
\]
where $\rCl{0,1}$ is the space generated by the vector $e_0$ and 
$\rCl{p,0}$ is the space generated by the Euclidean space orthogonal 
to $e_0$. A basis for $\rCl{0,1}$ has the $2^r$ elements:
\[		\fl
	\{e_{\ul{P}}\}=\{\ 1\ot\cdots\ot 1\ot 1\ ,
	\ 1\ot\cdots\ot 1\ot\;\! e_0\ ,\ 1\ot\cdots\ot\;\! e_0\ot 1\ ,
	\ \ \ldots\ \ ,\ e_0\ot\cdots\ot\;\! e_0\ot\;\! e_0\  \} .
\]
Now, we expand $A'$ in the basis $\{e_{\ul{P}}\}$ of $\rCl{0,1}$
with components in $\rCl{p,0}$
\[
	A'=A'^{\ul{P}}\ e_{\ul{P}}\qquad
	A'^{\ul{P}}\in\rCl{p,0}\subset\rCl{p,1} \qquad 
	e_{\ul{P}}\in\rCl{0,1}\subset\rCl{p,1} .
\]
Observe that both `components' and `basis', are elements of the algebra 
$\rCl{p,1}$, thus the product in $A'^{\ul{P}}\ e_{\ul{P}}$ is the $r$-fold
Clifford product. This finally completes the proof
\beqa*		\fl
	(-1)^r \prjs0{\ul{e_0}\, A'\, \ul{e_0}\,\conju{A'}}
	&=&\prjs0{\big(A'^{\ul{P}}\ e_{\ul{P}}\,\big)\,\ul{e_0}\, 
	\big(\conju{e_{\ul{Q}}}\,\conju{{A'}^{\ul{Q}}}\,\big)\, {\ul{e_0}}^{-1}}
	= \prjs0{\big({A'}^{\ul{P}}\ e_{\ul{P}}\,\big)\, 
	\big(\conju{e_{\ul{Q}}}\,\rev{{A'}^{\ul{Q}}}\,\big)}
		\\
	&=& \prj0{e_{\ul{P}}\conju{e_{\ul{Q}}}}
	\prjs0{{A'}^{\ul{P}}\rev{{A'}^{\ul{Q}}}\,}
	=\sum_{\ul{P}}\prjs0{{A'}^{\ul{P}}\ \rev{{A'}^{\ul{P}}}\,}\geq 0 .
\eeqa*
The last summation is always positive since, $\rCl{p,0}$ being an algebra
generated by an Euclidean metric, $\prj0{B\rev{B}\,}$ is a positive defined 
norm $\forall B\in\rCl{p,0}$. }}

Besides, the equality is satisfied if and only if every
$A'^{\ul{P}}=0$, that is, iff $A'=0$:
\[ 
	\prj0{A \ \ul{u}\ \conju{A}\ \ul{v}\,}=0 \quad\Leftrightarrow\quad
	A'=\conju{R_{\ul{u}}} A R_{\ul{v}}=0 .
\]
If we consider only time-like vectors $u_i$ and $v_i$,
then the operators $R_{\ul{u}}$ and $R_{\ul{v}}$ are
invertible. In this case $A'=0$ implies that
$A={\conju{R_{\ul{u}\!\!}}\,}^{-1}A' R_{\ul{v}}^{-1}=0$. This proves the 
result:

\theore{	\label{the:VanishTimeLike}
Let $A\in\rL$ be any $r$-fold multivector. There exists 
some set $\{u_i,v_i\}$ of $2r$ time-like vectors satisfying
\[
	T\{A\}_{\mu_1 \nu_1 \cdots \mu_r \nu_r}
	u_1^{\mu_1}v_1^{\nu_1}\cdots u_r^{\mu_r}v_r^{\nu_r} = 0 
\]
iff $A=0$. }

The general result including null vectors is the following:

\theore{	\label{the:VanishNull}
Let $A\in\rL$ be any $r$-fold multivector. There exists some 
set $\{u_i,v_i\}$ of $2r$ causal vectors satisfying
$
	T\{A\}_{\mu_1 \nu_1 \cdots \mu_r \nu_r}
	u_1^{\mu_1}v_1^{\nu_1}\cdots u_r^{\mu_r}v_r^{\nu_r} = 0
$
\ iff \ $\ul\ell\, A\,\ul k=0$, where
\beqa*
	\ul\ell=\bigotimes_{i=1}^r \ell_i & \qquad \mbox{with} \quad &
	\ell_i= \left\{ \begin{array}{ll}
	1 &\mbox{if $u_i$ is time-like}\\
	u_i \  &\mbox{if $u_i$ is null}  \end{array} \right.
\\
	\ul k=\bigotimes_{i=1}^r k_i &\qquad \mbox{with} \quad &
	k_i= \left\{ \begin{array}{ll}
	1 &\mbox{if $v_i$ is time-like}\\
	v_i \ &\mbox{if $v_i$ is null.}  \end{array} \right.
\eeqa*
}

\prove{First, we take into account that any null vector $\ell$ satisfies 
$\ell=\l \ell e_0 \conju \ell$, with $e_0$ a time-like vector and $\l\in\R^*$. 
This enables us to define a new $A'=\conju{\ul\ell}\, A\,\ul k$, 
following a manipulation analogous to (\ref{Aprime}).
Then, there appears the tensor $T\{A'\}$ 
contracted with $2r$ time-like vectors, since every null vector has been
replaced by $e_0$. Hence, from theorem \ref{the:VanishTimeLike}, the 
necessary and sufficient condition is $A'=0$. }

\section{Principal null directions}
		\label{sec:PrinNulDir}

A good example which illustrates the possibilities of the $r$-fold Clifford
formulation is the treatment of the principal null directions of a tensor.
In special and general relativity, the concept of principal null directions 
has been mainly used in the algebraic classification of the electromagnetic 
field and Weyl tensor (Petrov classification)
\cite{Sachs1964,Penrose1986-2}, which are, respectively, a bivector and a 
double bivector.
Let us introduce a general definition for arbitrary $r$-fold multivectors.

\define{	\label{def:Principal}
A null direction, represented by a null vector $\ell$, is a {\em principal
null direction (PND)} of an $r$-fold multivector $A\in\rL$, if it satisfies the 
equation
\[
	\ul\ell\, A\, \ul\ell=0 \qquad{\rm where}\quad 
	\ul\ell=\bigotimes^r \ell .
\]
This is equivalent to $(\ul\ell\inn A) \w \ul\ell=0$.
Or written with tensorial indices for a degree-defined $r$-fold multivector 
\[
	\ell^{\mu_1}\ell_{[\rho_{11}}\, 
	A_{\rho_{12}\cdots\rho_{1 n_1}]\mu_1\ \cdots\cdots\ 
	\mu_r[\rho_{r2}\cdots\rho_{r n_r}}\ \ell_{\rho_{r1}]}\ell^{\mu_r}=0
\]
where in each of the $r$ blocks we have one contraction with $\ell$ and the 
antisymmetrization of the rest of the indices with $\ell$: 
$\ \ell^{\mu_i}\ell_{[\rho_{i1}}\, 
A^{\cdots}{}_{\rho_{i 2}\cdots\rho_{i n_i}]\mu_i}{}^{\cdots}$.
}

We must note that this definition implies some unexpected features: a 1-vector
$A\in \L_1$ has no PND if it is time-like, one PND if it is light-like or 
an infinite number of PNDs if it is space-like.

\theore{A null vector $\ell$ is a PND of $A\in\rL$ iff
\[
	T\{A\}_{\mu_1\nu_1\cdots\mu_r\nu_r} 
	\ell^{\mu_1}\ell^{\nu_1}\cdots \ell^{\mu_r}\ell^{\nu_r}=0  .
\]
}

\prove{It follows from theorem \ref{the:VanishNull} and definition 
\ref{def:Principal}. }

This theorem is a generalization of the well-known result \cite{Penrose1986-1} 
relating the PNDs of the Weyl tensor, 
$\ell^\mu\ell_{[\alpha}C_{\beta]\mu\nu[\gamma}\,\ell_{\delta]}\ell^\nu=0$, 
to the null directions that make the BR superenergy tensor vanish, 
$T\{C\}_{\mu\nu\rho\sigma}\ell^\mu\ell^\nu\ell^\rho\ell^\sigma=0$.

\section{Conserved energy and superenergy tensors}
		\label{sec:Conserved}

We have shown that the basic superenergy tensors $T\{A\}$ satisfy the DP 
independently of the generating tensor $A$. The other important property 
expected to be satisfied by the energy--momentum tensor of any isolated 
physical field (or the total energy--momentum) is its local 
conservation: $\nabla^{\mu}T_{\mu\nu}=0$. 
The extension for general superenergy tensors is simply 
\beq
		\label{Divergence}
	\nabla^{\mu_1}T_{\mu_1\nu_1\cdots\mu_r\nu_r}=0 .
\eeq
Obviously, this property depends on the differential equations governing the 
physical field, namely on the dynamics of the field.
In this section we give (theorem \ref{the:SufCond}) a sufficient condition for 
the generating field $A$,
which guarantees the conservation of its basic superenergy tensor $T\{A\}$.
Moreover, this sufficient condition is satisfied by some physical fields such 
as the electromagnetic field, the Klein--Gordon field and the Riemann and Weyl 
tensors in Einstein spaces. Consequently, the divergence-free 
property of the standard electromagnetic and Klein--Gordon energy--momentum 
tensors, and the Bel and Bel--Robinson superenergy tensors, is unified in a 
unique procedure. The case of the current of the Dirac field is also embraced 
by this condition, if we consider a generalization of it 
(theorem \ref{the:SufCond3}) which will require
some discussion. A characteristic feature of the sufficient condition 
presented here is that its simplest and most natural expression is written 
using Clifford algebras. 
However, we will also write its standard tensorial equivalent.

First, let us write equation (\ref{Divergence}) for the basic superenergy
tensor using the $r$-fold Clifford algebra formalism:
\[
	\nabla\inn T\{A\}(\ul{u})=
	(-1)^r\f12 
	\prj{0,1,\cdots,1}{\,\dot\nabla \dot A\ul{u}\dot{\conju{A}\,}} .
\] 
Here some notational conventions are used. The differential operator 
$\nabla\equiv e^\mu\nabla_\mu$ is algebraically a vector, hence it multiplies 
the first block.
The overdots are used to indicate which elements must be derived by the 
operator. In the absence of overdots, $\nabla$ affects only the element
found immediately to its right-hand side. Note that the set of vectors 
$\ul u$ must not be derived, since we want to derive only the tensor $T\{A\}$.
Using the Leibnitz rule and the cyclic property of the scalar part of a 
product, we obtain
\beq		\fl
		\label{NablaDivergence}
	\nabla\inn T\{A\}(\ul{u})=
	(-1)^r \f12 \big(\prj{0,1,\cdots,1}{\nabla\! A\,\ul{u}\,\conju{A}\,}-
	\prj{0,1,\cdots,1}{A\,\ul{u}\,\conj{\nabla\! A}\,} \big)
	=(-1)^r \prj{0,1,\cdots,1}{\nabla\! A\,\ul{u}\,\conju{A}\,} .
\eeq
Here we see that $\nabla\!A$ is the unique differentiated quantity which 
contributes to the divergence. As is evident, $\nabla\!A=0$ is a sufficient
condition for the superenergy tensor $T\{A\}$ being divergence-free. But 
there is a more general sufficient condition:

\theore{\label{the:SufCond}%
Let $A\in\rL$ be any $r$-fold multivector. If $A$ satisfies the
condition
\beq
		\label{SufficientCond}
	\nabla\!A=\l A\ ,\qquad{\rm with}\quad \l\in\R 
\eeq
then the superenergy tensor $T\{A\}$ is divergence-free. }

\prove{From (\ref{NablaDivergence}) we get
\[
	\nabla\inn T\{A\}(\ul{u})=
	(-1)^r \l \prj{0,1,\cdots,1}{A\,\ul{u}\,\conju{A}\,}=0\qquad
	\forall A\in\rL
\]
if we take into account the properties of the Clifford conjugation:
\[		\fl
	\prj{0,1,\cdots,1}{A\,\ul{u}\,\conju{A}\,}=
	(-1)^{r-1}\prj{0,1,\cdots,1}{\conj{A\,\ul{u}\,\conju{A}\,}}=
	(-1)^{r-1}\prj{0,1,\cdots,1}{A\,\conj{\ul{u}}\,\conju{A}\,}=
	-\prj{0,1,\cdots,1}{A\,\ul{u}\,\conju{A}\,} .
\] 
\vspace{-1.4\baselineskip} } \vspace{0.4\baselineskip}

In order to express condition (\ref{SufficientCond}) in index tensorial 
notation, we need to split $A$ into its different degrees for the first block:
\[
	A=\sum_k{A_{[k]}}\qquad{\rm where}\quad 
	A_{[k]}=\frac{1}{k!}
	A_{[\mu_1\cdots\mu_k]\uls{C}}\ e^{\mu_1\cdots\mu_k}\ot e^{\uls{C}}
	\,\in\L_k\ot\L\ot\cdots\ot\L
\]
and then collect the contributions to different degrees coming from exterior
and inner products:
\beq
		\label{nablaSplit}
	\nabla\!A=\l A\quad\Leftrightarrow\quad
	\nabla\w A_{[k-1]}+\nabla\inn A_{[k+1]}=\l A_{[k]} 
	\quad\forall k=0,\ldots,d
\eeq
where
\[		\fl
	(\nabla\w A_{[k-1]})_{\mu_1\mu_2\cdots\mu_k\uls{C}}=
	k\nabla_{[\mu_1} A_{\mu_2\cdots\mu_k]\uls{C}}
	\quad{\rm and}\quad
	(\nabla\inn A_{[k+1]})_{\mu_1\mu_2\cdots\mu_k\uls{C}}=
	\frac{1}{k+1}\nabla^{\nu} A_{\nu\mu_1\mu_2\cdots\mu_k\,\uls{C}} .
\]

This sufficient condition is satisfied by diverse physical fields, so that 
the vanishing of the divergence of their energy or superenergy tensors
follows from theorem \ref{the:SufCond}:
\begin{itemize}
\item The source-free electromagnetic field. The full set of Maxwell equations
for the Faraday bivector takes the form $\nabla F=0$,
and the standard energy--momentum tensor is $T\{F\}$. Then, it immediately 
follows: $\nabla\inn T\{F\}=0$.

\item The Klein--Gordon field:
\[
	\nabla^2\phi=m^2\phi .
\]
Its standard energy--momentum tensor is generated by $(\nabla+m)\phi\in\L$:
\beqa*		\fl
	T\{(\nabla+m)\phi\}_{\mu\nu}&=&
	T\{\nabla\!\phi\}_{\mu\nu}+T\{m\phi\}_{\mu\nu}=
	-\f12\prj0{e_\mu \nabla\!\phi \,e_\nu \conj{\nabla\!\phi}}
	-\f12\prj0{e_\mu m\phi\, e_\nu \conj{m\phi}}
		\\
	&=& \nabla_\mu\phi\nabla_\nu\phi
	-\f12 \nabla_\rho\phi\nabla^\rho\phi g_{\mu\nu}
	-\f12 m^2\phi^2 g_{\mu\nu} .
\eeqa*
We can easily check that it satisfies condition (\ref{SufficientCond}),
\beq
		\label{K-G-condition}
	\nabla(\nabla+m)\phi=m(\nabla+m)\phi\quad\Rightarrow\quad
	\nabla\inn T\{(\nabla+m)\phi\}=0 .
\eeq

\item The Weyl conformal and the Riemann tensor in Einstein spaces. 
Einstein spaces satisfy $\nabla R=\nabla C=0$. For $d=4$, these conditions 
easily follow from the differential Bianchy identity, $\nabla\w R=0$, and 
the Lanczos identity, $R=-*\!R*$ (where $*\!R*$ is the double Hodge 
dual of the Riemann). For $d>4$ the equations also hold, but they are a bit 
more difficult to show. 
Therefore, we obtain
\[
	\nabla C=\nabla R=0\quad\Rightarrow\quad \nabla\inn T\{C\}=
	\nabla\inn T\{R\}=0
\]
which says that the generalizations of the Bel--Robinson, $T\{C\}$, and the 
Bel, $T\{R\}$, superenergy tensors are divergence-free in Einstein spaces of 
any dimension.

\end{itemize}

\define{An {\em $r$-fold multivector differential operator} ${\cal D}\in\rL$ is
a differential operator which algebraically belongs to $\rL$. 
${\cal D}$ will be said to be {\em constant} if it is formed from any 
combination of constant multivectors and the differential operator $\nabla$. } 
Some examples of constant $2$-fold multivector differential operators are: 
$1\ot(\nabla+m)$, $\nabla\ot\nabla$ and
$B\ot(v\w\nabla)$, with $m,B,v\in\L$ constants. 

Now consider that we have an $r$-fold multivector $A\in\rL$ satisfying 
condition (\ref{SufficientCond}) with $\l$ constant. Let ${\cal D}\in\sL$
be a constant $s$-fold multivector differential operator.
We can then construct the $r'$-fold multivector 
\[
	A'\equiv\dot A\dot{\cal D}\ \in\,\mbox{\Large ${}^{r'}$}\!\!\L
	\qquad{\rm where}\quad r'=\max(r,s)
\] 
where ${\cal D}$ differentiates $A$. A simple example is provided by 
$A'=\dot A(1\ot\dot\nabla)=\dot A\ot\dot\nabla\equiv\nabla_{\mu}A\ot e^{\mu}$. 
Then, the $r'$-fold multivector $A'$ also satisfies
condition (\ref{SufficientCond}), except for terms involving the 
non-commutativity of derivatives: $\nabla_{[\mu}\nabla_{\nu]}$, i.e. involving 
terms linear in the Riemann tensor or its derivatives:
\beq
		\label{RiemannTerms}
	\nabla(\dot A\dot{\cal D})=\dot{(\nabla A)}\dot{\cal D}+{\cal O}(R)=
	\l(\dot A\dot{\cal D})+{\cal O}(R) .
\eeq
Hence, for flat spacetimes we get the result:
\theore{	\label{the:Construct}
Let $A\in\rL$ be any $r$-fold multivector on a flat spacetime $\cal M$,
satisfying 
$
	\nabla\!A=\l A\ $, with $\l$ constant. 
Then, for any constant $s$-fold multivector differential operator $\cal D$,
$A'=\dot A\dot{\cal D}$ also satisfies 
$
	\nabla\! A'= \l A'\ 
$.
}

\prove{Trivial from (\ref{RiemannTerms}).}

Thus, from theorem \ref{the:SufCond} the superenergy tensor $T\{A'\}$ 
will be divergence-free.

\theore{	\label{the:Construct0}
Let $A\in\rL$ be any $r$-fold multivector on a flat spacetime $\cal M$,
satisfying $\nabla\!A=0$, then the ($r$$+$$1$)-fold multivector 
$A'=\nabla\ot A$ also satisfies $\nabla\!A'=0$. }

\prove{On a flat spacetime, the Clifford square of the operator nabla is 
algebraically a scalar $\nabla^2\in\L_0$. Thus, it easily follows that
\[
	\nabla A'=\nabla^2\ot A=1\ot\nabla^2 A=1\ot \nabla(\nabla A)=0 .
\]
\vspace{-1.2\baselineskip} } \vspace{0.2\baselineskip}

Thus, from theorem \ref{the:SufCond}, the derived tensors $A'$ obtained in 
any of these two theorems, will generate a divergence-free superenergy tensor 
$T\{A'\}$.
Examples of these constructions are provided by the 
superenergy tensors proposed by Senovilla \cite{Seno1999b} and previously 
introduced in \cite{Chevreton1964,Teyssandier1999}, for the 
electromagnetic and Klein--Gordon fields:

\begin{itemize}

\item For the source-free electromagnetic field, the rank-4 
superenergy tensor which has been considered \cite{Chevreton1964,Seno1999b}
is the full symmetrization of $T\{\nabla\ot F\}$. 
From theorems \ref{the:Construct0} and \ref{the:SufCond}, it follows that the 
tensor $T\{\nabla\ot F\}$ is divergence-free for flat spacetimes:
\[
	\nabla F=0\quad\Rightarrow\quad
	\nabla(\nabla\ot F)=0 \quad\Rightarrow\quad 
	\nabla\inn T\{\nabla\ot F\}=0 .
\]
However, there the divergence of the tensor $T\{\dot F\ot\dot\nabla\}$
also contributes in the divergence of the symmetrized tensor.
This term also vanishes in flat spacetimes, as
follows from theorems \ref{the:Construct} and \ref{the:SufCond}: 
\[
	\nabla F=0 \quad\Rightarrow\quad
	\nabla(\dot F\ot \dot\nabla)=0\quad\Rightarrow\quad
	\nabla\inn T\{\dot F\ot \dot\nabla\}=0 .
\]
Thus, the symmetrized electromagnetic superenergy tensor is divergence-free.

\item For the Klein--Gordon field the rank-4 superenergy 
tensor \cite{Teyssandier1999,Seno1999b}
is $T\{(\nabla+m)\ot(\nabla+m)\phi\}$. From (\ref{K-G-condition}) we
know that $(\nabla+m)\phi$ satisfies condition (\ref{SufficientCond}) with
$\l=m$ constant. Therefore, applying theorem \ref{the:Construct}, for flat
spacetimes we get
\beqa*
	\nabla((\nabla+m)\ot(\nabla+m)\phi)=m((\nabla+m)\ot(\nabla+m)\phi)
	\\
	\Rightarrow\quad
	\nabla\inn T\{(\nabla+m)\ot(\nabla+m)\phi\}=0 .
\eeqa*

\end{itemize}

Let us now introduce a generalization of the sufficient condition 
(\ref{SufficientCond}), which we will apply to the current of the Dirac 
field. We will generalize (\ref{SufficientCond}) in two steps. The first
consists in allowing the scalar $\l$ to be a more general multivector.

\theore{\label{the:SufCond2}%
Let $A\in\rL$ be any $r$-fold multivector. If $A$ satisfies the
condition
\beq
		\label{SufficientCond2}
	\nabla\!A=B A\qquad{\rm with}\quad B\in\L \quad{\rm and}\quad
	B-\conju{B}=0
\eeq
then the superenergy tensor $T\{A\}$ is divergence-free. }

\prove{Taking expression (\ref{NablaDivergence}) we get
\[		\fl
	(-1)^r 2\nabla\inn T\{A\}(\ul{u})=
	\prj{0,1,\cdots,1}{BA\,\ul{u}\,\conju{A}\,}-
	\prj{0,1,\cdots,1}{A\,\ul{u}\,\conju{A}\conju{B}\,}=
	\prj{0,1,\cdots,1}{(B-\conju{B})A\,\ul{u}\,\conju{A}\,} .
\]
\vspace{-1.4\baselineskip} } \vspace{0.4\baselineskip}

\no Observe that the result of theorem \ref{the:Construct} is trivially 
generalized for the condition (\ref{SufficientCond2}).

The second step consists in allowing factors multiplying $A$ on both sides 
to appear: $\nabla A=BAC$. But in this case, the divergence of the superenergy tensor $T\{A\}$ depends on the commutativity of the right-hand factor $C$ with 
the vectors to which $T\{A\}$ is applied. Hence, the result is not the vanishing
of the divergence, but of the divergence applied to some collection of $r$ 
vectors: $\nabla\inn T\{A\}(\ul{u})=0$. For this reason, the result can be 
applied to the current of a Dirac field $j=-2T\{\Psi\}(e_0)$.

\theore{\label{the:SufCond3}%
Let $A\in\rL$ be any $r$-fold multivector and $\ul{u}\in\bigotimes^r \L_1$ any
$r$-fold vector. If they satisfy the
conditions
\beqna
		\label{SufficientCond3}
	\nabla\!A=\sum_i B_i A C_i \qquad{\rm with}\quad 
	B_i\in\L\,,\ C_i\in\rL \\
	\nonumber
	{\rm and}\qquad B_i=s_i\conju{B}\!_i\qquad  
	C_i\ul{u}=s_i\ul{u}\conju{C}\!_i
	\qquad{\rm where}\quad s_i=\pm1
\eeqna
then the divergence of the superenergy tensor contracted with $\ul{u}$ vanishes:
$\nabla\inn T\{A\}(\ul{u})=0$. }

\prove{Taking expression (\ref{NablaDivergence}) we get
\beqa*
        (-1)^r 2\nabla\inn T\{A\}(\ul{u})&=&
        \sum_i \prj{0,1,\cdots,1}{B_iAC_i\,\ul{u}\,\conju{A}\,}-
        \prj{0,1,\cdots,1}{A\,\ul{u}\,\conju{C}\!_i\conju{A}\conju{B}\!_i\,}
		\\
        &=&\sum_i \prj{0,1,\cdots,1}{B_iA(C_i\,\ul{u})\,\conju{A}\,}-
        \prj{0,1,\cdots,1}{\conju{B}\!_iA\,(\ul{u}\,\conju{C}\!_i)\conju{A}\,}
	.
\eeqa*
\vspace{-2.4\baselineskip} \\ \mbox{} } \vspace{0.4\baselineskip}

\corollar{If $\ul{u}$ is constant then $T\{A\}(\ul{u})$ is divergence-free. }
\corollar{If $T\{A\}$ is completely symmetric:
$
	T\{A\}_{\mu_1\nu_1\cdots\mu_r\nu_r}=
	T\{A\}_{(\mu_1\nu_1\cdots\mu_r\nu_r)}\ 
$,
and $\ul u$ is a collection of Killing vectors, then $T\{A\}(\ul{u})$ is 
divergence-free. }

The Hestenes' real multivector formulation of the Dirac equation for a spinor
minimally coupled with an electromagnetic field is 
\cite{Hestenes1973}
\[
	\nabla \Psi=m\Psi e_{012}+ q A\Psi e_{12}
\]
where $\Psi\in\Cl{3,1}^+$ is the operator spinor described in section 
\ref{sec:Lorentz&Spinors}, and $A$ is the electromagnetic vector potential.
Observe that, although the {\it Dirac--Hestenes 
equation} is usually written for the signature $1-3$, it can also be formulated 
for the signature $3-1$ \cite{Parra1992}. We can easily see that $\Psi$ 
satisfies the conditions of the theorem \ref{the:SufCond3}, with 
$B_1=m$, $C_1=e_{012}$, $B_2=qA$ and $C_2=e_{12}$, for $j=-2T\{\Psi\}(e_0)$:
\[
	\left.  \begin{array}{c} 
	m=\conj{m}\\ e_{012}\,e_0=e_0\,\conj{e_{012}} 
	\end{array} \right\} 
	\quad{\rm and}\quad 
	\left.  \begin{array}{c} 
	qA=-\conj{qA}\\ e_{12}\,e_0=-e_0\,\conj{e_{12}} 
	\end{array} \right\} 
	\quad\Rightarrow\quad
	\nabla\inn T\{\Psi\}(e_0)=0 .
\]
Since $e_0$ is an inertial observer in special relativity, the current is
conserved:
\[
	\nabla\inn j=0 .
\]

\vspace*{-2pt}
\section*{Acknowledgments}
The authors gratefully acknowledge J M Senovilla for his careful reading 
of the manuscript and several suggestions, and the referees for their useful 
comments. This work has been supported by the scholarship AP96-52209390
and the contracts BFM2000-0604 from the Spanish Ministry of Education,
and 2000SGR/23 from the DGR of the Generalitat de Catalunya.

\end{document}